\documentclass[epj,twocolumn]{webofc}
\usepackage[varg]{txfonts}   
\usepackage{subfig}
\usepackage{siunitx}
\woctitle{Powders \& Grains 2017}
\begin{document}
\title{Dynamics of wetting explored with inkjet printing}
%
%

\author{\firstname{Simeon} \lastname{Völkel}\inst{1}\fnsep\thanks{\email{simeon.voelkel@uni-bayreuth.de}} \and
        \firstname{Kai} \lastname{Huang}\inst{1}\fnsep\thanks{\email{kai.huang@uni-bayreuth.de}}
}

\institute{Experimentalphysik V, University of Bayreuth, 95440 Bayreuth, Germany
          }

\abstract{%
An inkjet printer head, 
which is capable of depositing liquid droplets with a resolution of $22$ picoliters and high repeatability, 
is employed to investigate the wetting dynamics of drops printed on a horizontal plane 
as well as on a granular monolayer. 
For a sessile drop on a horizontal plane, 
we characterize the contact angle hysteresis, drop volume and contact line dynamics 
from side view images. 
We show that the evaporation rate scales with the dimension of the contact line 
instead of the surface area of the drop. 
We demonstrate that the system evolves into a closed cycle 
upon repeating the depositing-evaporating process, 
owing to the high repeatability of the printing facility. 
Finally, 
we extend the investigation to a granular monolayer 
in order to explore the interplay between liquid deposition and granular particles.
}
\maketitle
\section{Introduction}
\label{chap:intro}

Granular materials are ubiquitous in nature, industry and daily lives~\cite{Duran00}. 
Despite intense investigations motivated by applications 
in geoscience, chemical and civil engineering, 
the physics of granular materials, 
are still far from being understood,
particularly when they are partially wet~\cite{Herminghaus2005,Mitarai2006,Huang2014}. 
The mechanical properties of a granular material 
change dramatically as a tiny amount wetting liquid is added, 
representing the formation of capillary bridges between adjacent particles~\cite{Halsey1998,Huang2009a}. 
The enhancement of rigidity leads to an easily moldable material 
such as wet sand on the beach for sculpturing~\cite{Scheel2008,Pakpour2012}. 
In the pendular regime~\cite{Mitarai2006} with capillary bridges formed between adjacent particles, 
previous investigations showed that the collective behaviors of partially wet granular materials, 
such as clustering, phase transitions and pattern formation, 
are related to the capillary force induced by the wetting liquid~\cite{Huang2009b,Huang2011,Huang2012,May2013,Butzhammer2015}. 
However, 
as more liquid is added, 
it is still challenging to understand how a wetting liquid film advances 
inside a geometrically heterogeneous granular packing, 
as well as how the particles respond to the capillary interactions 
induced by the additional wetting liquid.

Wetting 
is also an ubiquitous phenomenon that has attracted interest 
from physics, chemistry and engineering communities over the past decades~\cite{Gennes1985,Bonn2009}. 
More recently, 
there has been a growing interest in understanding the wetting dynamics 
on patterned and heterogeneous surfaces, 
motivated by a better control of wettability and liquid transport~\cite{Semprebon2014b,Sbragaglia2014}. 
Nevertheless, 
there are still open questions 
(e.\,g. contact angle hysteresis, contact line dynamics at the onset of the pinning-depinning transition, etc.)
\cite{Thiele2006,Eral2013,semprebon2014}. 

Here, 
we use a commercial inkjet printer head (HP~51645A) 
to explore the dynamics of wetting, 
because it enables a fine control of the drop volume with high repeatability. 
For both a horizontal plane and a granular monolayer, 
repeating the print-evaporate process leads to reproducible contact line dynamics.

\section{Setup}
\label{chap:setup}

\begin{figure}[b]
\centering
\includegraphics[width=0.35\textwidth,clip]{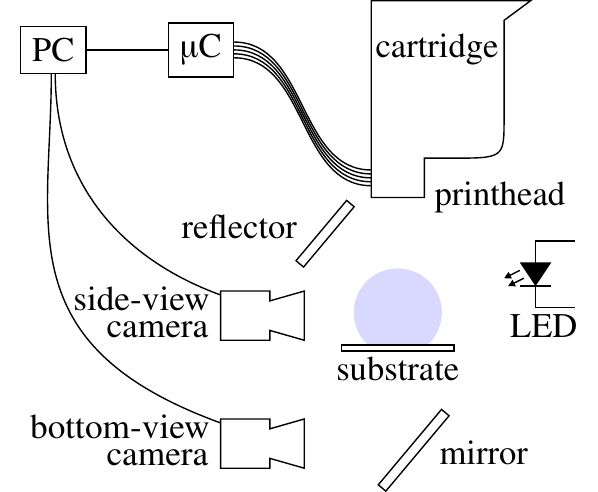}
\caption{%
Schematic diagram of the experimental setup. 
$\mu$C stands for microcontroller. 
A reflector is used to provide back-light illumination for the bottom view camera.
}
\label{fg:setup}       
\end{figure}

Figure~\ref{fg:setup} shows a sketch of the experimental setup.
An empty print catridge (HP~C6125A) is filled with purified water
(LaborStar TWF, surface tension $\gamma = \SI{0.072}{\newton\per\meter}$).
Its printhead is pointing downwards 
in order to deposit droplets vertically onto the horizontal substrate 
[polytetrafluoroethylene (PTFE), and a monolayer of glass beads (SiLiBeads Type S)]. 
In order to control all nozzles (printing frequency \SI{2.8}{\kilo\hertz}) individually, 
we use a custom microcontroller board (Microchip PIC18F6722).
Two types of cameras 
(Lumenera LU135M, Nikon D7000) 
are used to take side view and bottom view images.

A computer program (PC) triggers the image acquisition for one experimental cycle 
$N_\text{print}$ times during the printing phase,
each time after a specific volume $V_\text{step}$ has been deposited,
and $N_\text{evap}$ times during evaporation phase, at a fixed interval $T_\text{evap}$.
The ambient temperature is regulated to \SI{20}{\celsius}.
The relative humidity is in the range from \SIrange{20}{40}{\percent}.
The average droplet volume is determined to be \SI{22.4+-1.1}{\pico\liter} by image analysis of side view images.

\section{Image analysis}
\label{chap:analysis}

\begin{figure}[b]
\centering
\subfloat[]{\label{fg:parameter}%
\includegraphics[width=0.23\textwidth,clip]{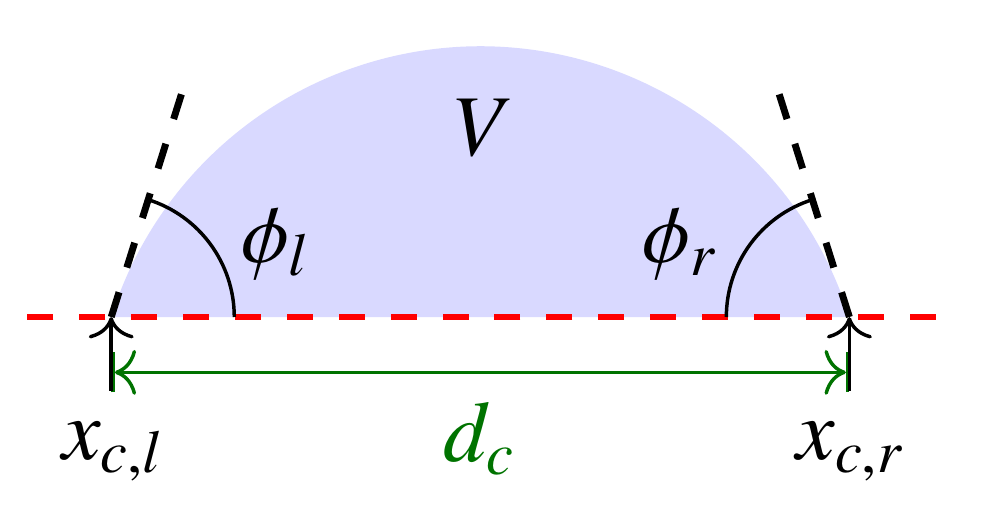}}
\subfloat[]{\label{fg:volume}%
\includegraphics[width=0.23\textwidth,clip]{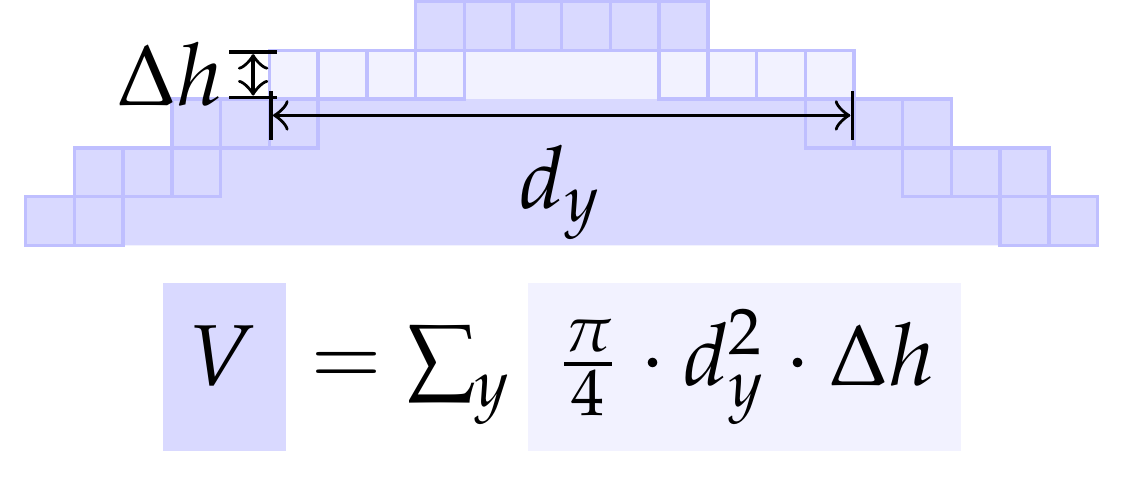}}

\subfloat[]{\label{fg:contoursegment}%
\includegraphics[width=0.23\textwidth,clip]{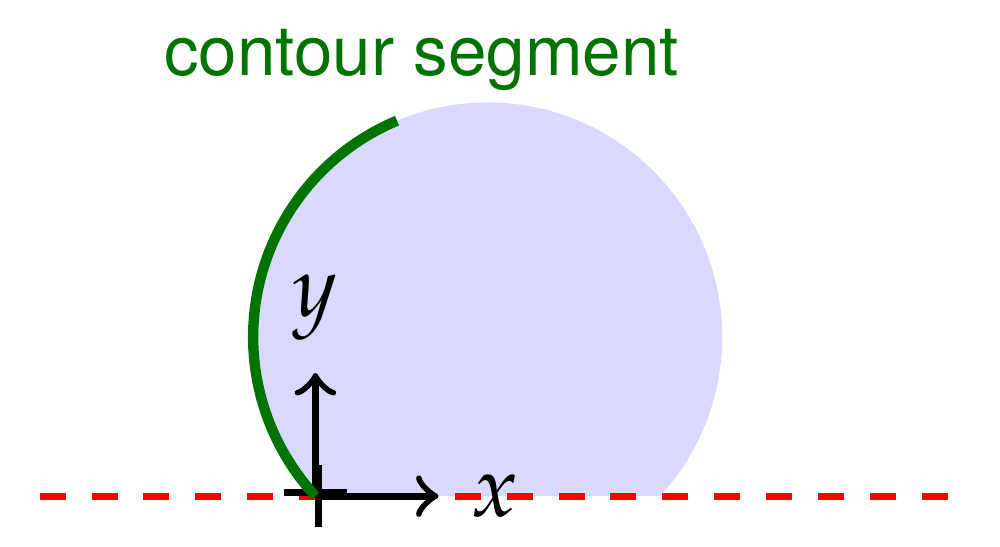}}
\subfloat[]{\label{fg:contourvektor}%
\includegraphics[width=0.23\textwidth,clip]{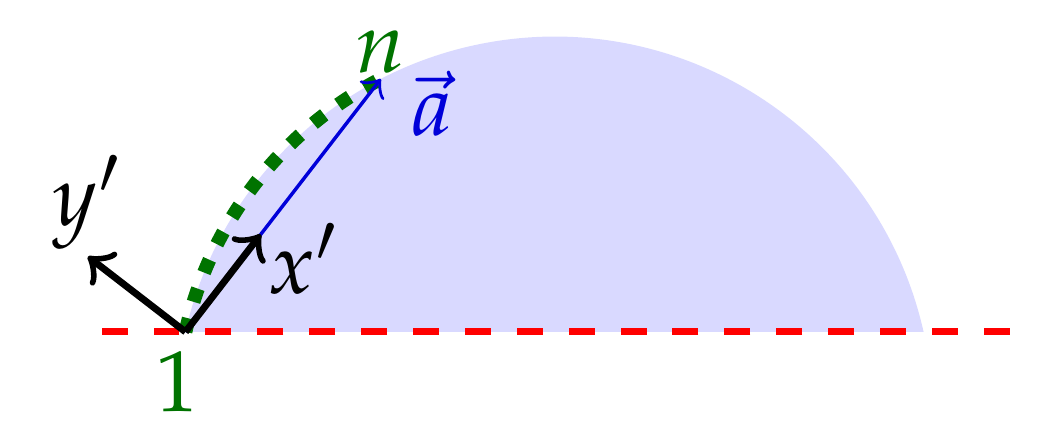}}
\caption{%
Image analysis procedure and definitions of drop quantities.
The red dashed lines indicate the substrate.
}
\label{fg:analysis}       
\end{figure}

Figure~\ref{fg:analysis} summarizes the image analysis procedure 
to obtain the geometric properties of the sessile drop.
The side view images are cropped such 
that the bottom of the rectangular region of interest (ROI) 
separates the drop from the substrate,
and binarized to segment the image.

Figure~\ref{fg:parameter} defines the quantities of the drop.
The volume $V$ is estimated assuming rotational symmetry along the vertical axis of the drop
in each line of the image,
as shown in Fig.~\ref{fg:volume}.
The left and right contact angles~$\phi_l$ and $\phi_r$ 
as well as the (horizontal) $x$-coordinates of the associated contact points $x_{c,l}$, $x_{c,r}$ 
are determined with subpixel resolution
by fitting locally a parabola using the standard least square method to the drop contour.
Short%
\footnote{For estimating the (vertical) radius of curvature of a sessile drop
close to the contact points,
its contour in the side view can be approximated as a part of an ellipse.
The short main axis gives a good estimate independent of the drop volume and contact angle.
Here, a contour segment one order of magnitude shorter is used.
Fig.~\ref{fg:contoursegment} is not to scale.},
adjacent segments of the contour are used, 
as indicated by Fig.~\ref{fg:contoursegment}.
As shown in Fig.~\ref{fg:contourvektor},
the fit is not performed in the laboratory frame $(x,y)$,
but in the rotated $(x',y')$ coordinate system 
set by the first ($1$) and last point ($n$) of the contour segment.
This is to circumvent ambiguity problems with contact angles close to \ang{90}
as sketched in Fig.~\ref{fg:contoursegment}
and increase both robustness and accuracy.
The diameter of the contact line~$d_c$ 
is estimated as the distance between the two contact points 
(cf.\ Fig.~\ref{fg:parameter}).

\section{Results and discussion}
\label{chap:results}

\begin{figure}[b]
\centering
\includegraphics[width=0.5\textwidth,clip]{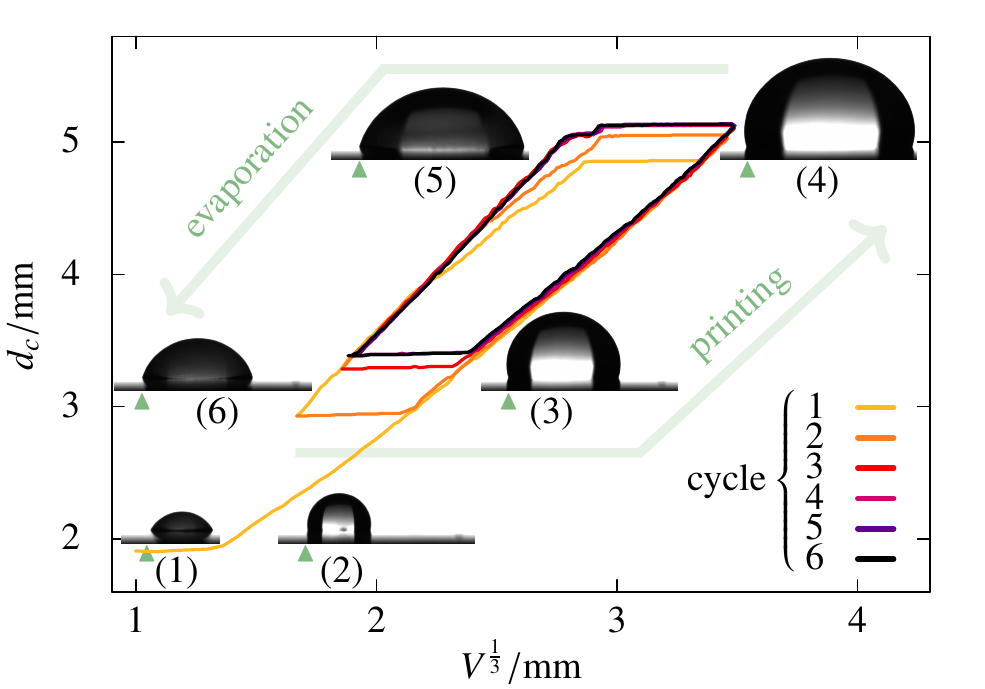}
\caption{%
Scaling of the contact line diameter $d_c$ with the drop volume $V$ for water on PTFE.
Six cycles of $N_\text{print}=\num{120}$, $V_\text{step}=\SI{0.34+-0.02}{\micro\liter}$, $N_\text{evap}=\num{120}$, $T_\text{evap}=\SI{60}{\second}$ are shown.
The triangles mark the same position on the substrate.
See text for details.
}
\label{fg:shape}       
\end{figure}

Figure~\ref{fg:shape} shows 
the development of $d_c$ 
as a function of $V^\frac{1}{3}$ 
for six print-evaporate cycles.
Starting with an initial drop (1),
its volume is increased 
by adding water droplets from above using the inkjet printhead.
At first, $d_c$ remains constant 
until the advancing contact angle is reached (2).
As further liquid is added, 
the drop expands laterally (3),~(4).
During evaporation, 
the drop again first keeps $d_c$ constant 
until the receding contact angle is reached (5).
Further evaporation lets the drop shrink laterally (6).
When adding some liquid again, 
the same behaviour recurs.
The overlapping curves for cycles \mbox{4--6} illustrate 
the excellent repeatability of the experimental setup
and 
that the deposition-evaporation process can be driven into a closed cycle 
by just periodically adding a fixed volume with high precision.

In the limit of small drops 
(i.\,e.\ small Bond-number $\mathrm{Bo} = \rho \, g \, V^\frac{2}{3} \, \gamma^{-1}$,
where $\rho$: fluid density, $g$: gravitational acceleration), 
interfacial tension determines the equilibrium shape of a drop
and
all linear dimensions of the drop scale with $V^\frac{1}{3}$. 
With growing volume, 
the relative influence of gravity increases 
and finally flattens the drop \cite{Extrand2010}.
The drops (2), (3) and (4) as well as (1), (5) and (6) have consistent shapes.
It is worth noting that both 
lateral expansion and shrinking of the drop 
follow a sloped straight line in Fig.~\ref{fg:shape}:
The sessile drops ($\mathrm{Bo} < \num{1.65}$) are thus small.
However, due to contact angle hysteresis, 
the shapes of an expanding and contracting drop differ,
which results in two distinct slopes in Fig.~\ref{fg:shape}.

\begin{figure}
\centering
\includegraphics[width=0.5\textwidth,clip]{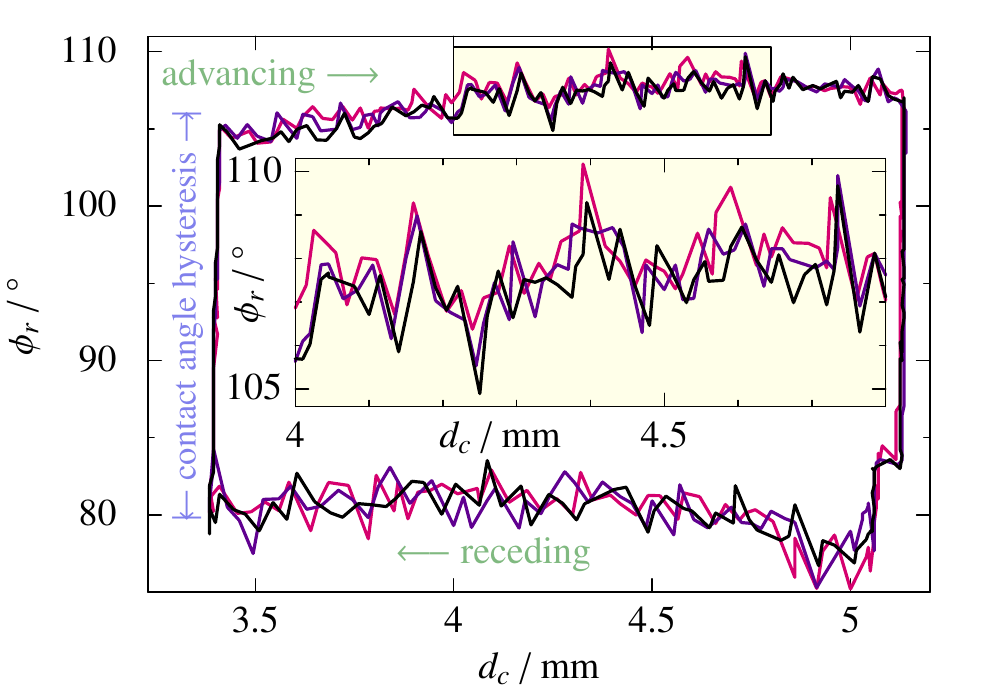}
\caption{Contact angle $\phi_r$ of water on PTFE measured from side view images at the freely moving edge.
Coloring corresponds to Fig.~\ref{fg:shape}.
}
\label{fg:cah}       
\end{figure}

\begin{figure}
\includegraphics[width=0.48\textwidth,clip]{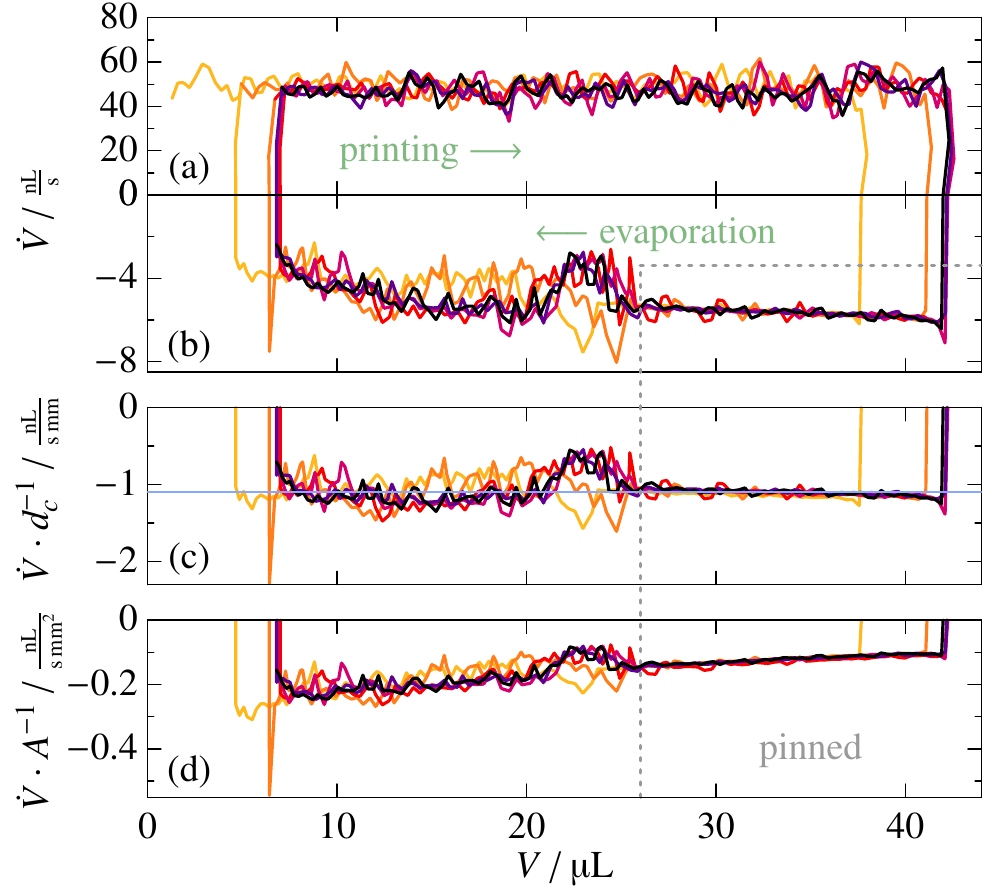}
\caption{(a) and (b): Effective printing and evaporation rate estimated from side view images.
(c) and (d): Comparison of two scaling laws for the evaporation rate.
The blue line corresponds to \SI{-1.1}{\nano\liter\per\second\per\milli\meter} 
and is a guide to the eye.
See text for details.
Coloring corresponds to Fig.~\ref{fg:shape}.}
\label{fg:evaporation}       
\end{figure}

Figure~\ref{fg:cah} shows the contact angle at the right edge $\phi_r$ 
plotted over $d_c$
during the last three print-evaporate cycles.
While the contact line is advancing or receding,
the contact angle remains approximately constant.
As the contact angle for the two directions of movement differs,
contact angle hysteresis is clearly visible.
In the inset images (1)--(6) of Fig.~\ref{fg:shape},
the drop expands to his right side, 
keeping his left edge at a fixed position.
This is due to inhomogenities on the substrate, 
resulting in a slightly larger contact angle hysteresis there 
than for the rest of the contact line.
Thanks to the pinned left side in the experiments shown in Fig.~\ref{fg:cah},
equal $d_c$ results in coincident positions $x_{c,r}$ of the right edge of the drop.
The inset in Fig.~\ref{fg:cah} magnifies and highlights a part of the graph.
The strong correlation between subsequent runs reveals 
that the fluctuations in the measured contact angles 
come from the drop's exploration of the surface heterogenities.
Figure~\ref{fg:evaporation} shows 
that the average deposition rate of the used printing-protocol 
exceeds the evaporation rate by an order of magnitude.
While evaporation drives the recedence of the contact line, 
its influence on the advance is marginal.
As a result, 
the advancing contact angle in Fig.~\ref{fg:cah} exhibits a more prominent correlation
than the measurements during the receding phase.

As the blue horizontal line in Fig.~\ref{fg:evaporation}c illustrates,
the evaporation rate scales with the contact line diameter.
Thus we conclude 
that the drop evaporates mainly through its three phase line
(see \cite{Hu2002}) 
instead of its surface area
(see Fig.~\ref{fg:evaporation}d for a comparison).
The scaling of the evaporation with the drop geometry reveals 
that our sessile drop is also small in this regard.

Having illustrated the repeatability provided by the setup
when using a flat substrate, 
we now extend our investigation to a granular monolayer.
Concerning wetting in a granular system, 
one of the most important challenges is to understand
the interplay between wetting liquid and granules.
The particles
are subject to capillary forces
if they are not immersed in liquid.
In a loose packing, 
a movement of the liquid interface 
can therefore lead to a movement of granules, 
if they are not fully constrained by their neighbours.
The details of the advancing and receding of the liquid interface, 
the involved wetting and dewetting, as well as rupture events of capillary bridges
are strongly dependend on the 
the configuration of neighboring particles \cite{Semprebon2016}.

For the following experiments, 
spherical glass beads  
of \SIrange{0.3}{0.4}{\milli\meter} diameter
are put on top of a horizontal glass slide.
By tapping they are brought into a two-dimensional random packing.
Further beads are added 
until the bottom of a $\num{25}\times\SI{25.5}{\milli\meter}$ rectangular enclosure is filled. 
After initial reorganization of some particles,
the system again can apparently be driven into a closed cycle
by periodically adding and evaporating water.

\begin{figure}
\centering
\subfloat[$V=\SI{0.56+-0.03}{\micro\liter}$]{%
\includegraphics[width=0.23\textwidth,clip]{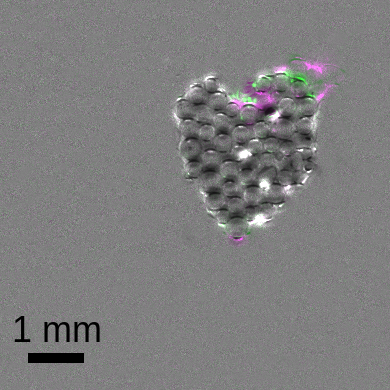}%
\label{fg:sandA}%
}\hspace*{1em}%
\subfloat[$V=\SI{1.22+-0.06}{\micro\liter}$]{%
\includegraphics[width=0.23\textwidth,clip]{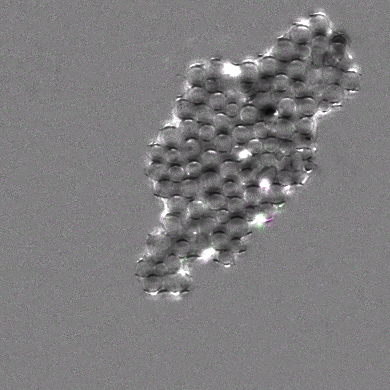}%
\label{fg:sandB}%
}\vspace{0.0em}

\subfloat[$V=\SI{1.68+-0.09}{\micro\liter}$]{%
\includegraphics[width=0.23\textwidth,clip]{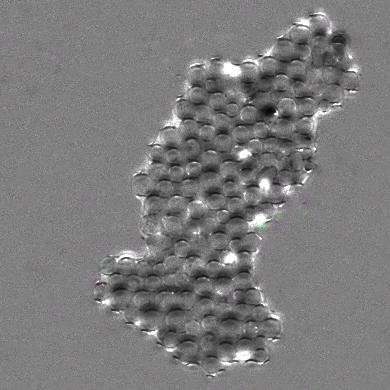}%
\label{fg:sandC}%
}\hspace*{1em}%
\subfloat[$V=\SI{2.24+-0.11}{\micro\liter}$]{%
\includegraphics[width=0.23\textwidth,clip]{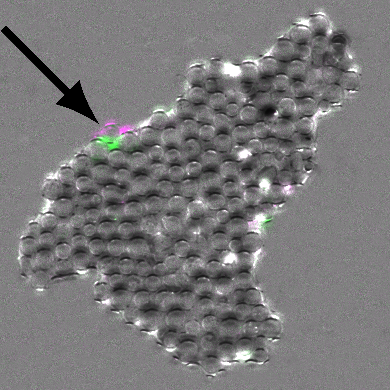}%
\label{fg:sandD}%
}
\caption{Overlay of typical bottom view images taken in two subsequent printing cycles
($N_\text{print}=\num{100}$, $V_\text{step}=\SI{22.4+-1.1}{\nano\liter}$, $N_\text{evap}=\num{40}$, $T_\text{evap}=\SI{30}{\second}$).
Unsaturated pixels show regions of identical wetting
while green and purple regions highlight the differences of the wetted area.
See, for example, the region marked by the arrow in (d).
}
\label{fg:sand}       
\end{figure}

Figure~\ref{fg:sand} compares two subsequent printing-cycles
on this three-dimensionally patterned substrate.
After removing the background by division, 
the overlaid images are obtained 
by putting into their red and blue channels 
the image of one cycle
and in their green channels the image of the subsequent one.
Qualitatively, Fig.~\ref{fg:sand} provides the following features: 
(i) The contour of a sessile drop on a granular layer
does not exhibit a circular contact line
due to the prominent structure of the substrate.
(ii) The deviation from a circle 
is more than the particle size,
especially if the wettability of the substrate is high.
This is partially due to the fact that
the wetting of a neighbouring particle is a discrete event
which leads to an abrupt change of the liquid interface position.
(iii) While for drops on an inclined homogeneous plane
the aspect ratio of the bottom view is a suitable parameter 
for describing, 
e.\,g., the onset of sliding \cite{semprebon2014},
the irregularly jagged contact line shown here raises the question:
Which shape descriptors are suitable for the case of a partially wetted granular layer?
Finally, 
the tiny difference in the shape of the wetted area 
(see colored regions in Fig.~\ref{fg:sand}) 
clearly illustrates that, 
even in this complex system, 
this setup offers a high repeability.

\section{Summary}
\label{chap:summary}

In this work, 
we use a commercial inkjet printer head, 
which provides a fine volume control of liquid deposition 
in a well controlled and cost-effective way, 
to explore the wetting dynamics of liquid drops 
on a plane and a granular monolayer.

For a sessile drop on a plane, 
we demonstrate the excellent repeatability of the setup 
through applying print-evaporate cycles 
and monitoring the drop geometry.
Contact angle hyteresis is observed.
The evaporation rate is found to scale with the diameter of the drop contact line,
which is in agreement with a previous investigation~\cite{Hu2002}.
Thanks to the high and adjustable printing rate (\SI{63+-3}{\nano\liter\per\second} per nozzle), 
water evaporation can be easily compensated 
by printing in typical experimental conditions.

Remarkably, 
we find that, 
even on a discrete substrate like a granular layer, 
this print-evaporate protocol leads to reproducible contour line dynamics. 
This feature facilitates further investigations 
on the interactions between a wetting liquid and a granular bed, 
in order to shed light on the advance of additive manufacturing 
as well as other drop-on-demand applications~\cite{Yarin2006,Derby2010}.

\end{document}